\documentstyle[prl,aps,epsf,floats,twocolumn,graphicx,amssymb]{revtex} 
\newcommand{\be}{\begin{equation}}
\newcommand{\ee}{\end{equation}}
\newcommand{\bea}{\begin{eqnarray}}
\newcommand{\eea}{\end{eqnarray}}

\begin{document}
\draft
\twocolumn[\hsize\textwidth\columnwidth\hsize\csname
@twocolumnfalse\endcsname

\title{Generalized Bose-Fermi statistics and structural correlations in weighted networks}
\author{Diego Garlaschelli$^{1}$, Maria I. Loffredo$^{2}$}
\address{$^1$Dipartimento di Fisica, Universit\`a di Siena, Via Roma 56, 53100 Siena ITALY\\
$^2$Dipartimento di Scienze Matematiche ed Informatiche, Universit\`a di Siena, Pian dei Mantellini 44, 53100 Siena ITALY}
\date{\today}
\maketitle
\begin{abstract}
We derive a class of generalized statistics, unifying the Bose and Fermi ones, that describe any system where the first-occupation energies or probabilities are different from subsequent ones, as in presence of thresholds, saturation, or aging. The statistics completely describe the structural correlations of weighted networks, which turn out to be stronger than expected and to determine significant topological biases. Our results show that the null behavior of weighted networks is different from what previously believed, and that a systematic redefinition of weighted properties is necessary.
\end{abstract}

]
\narrowtext
The Fermi-Dirac and Bose-Einstein distributions describe systems whose states can be discretely populated, at most once or an infinite number of times respectively. Even if they were originally introduced to model quantum particles, they turn out to describe a wider range of systems, including traffic \cite{quantumtraffic} and complex networks \cite{quantumginestra,newman_expo}. 
It is therefore not surprising that extensions of these distributions are indicated not only by quantum theory itself (e.g. anyons and supersymmetry), but also by other research fields where they are encountered \cite{quantumginestra}.
In this Letter, starting from a problem arising in network theory, 
we derive a class of generalized statistics that unify the Bose-Einstein and Fermi-Dirac ones by extending them in two directions simultaneously: first, the maximum occupation number of a state is any integer between one and infinity; second, the first-occupation energies may be different from next-occupation ones. 
A natural application is to social networks, where establishing a link representing mutual acquaintance between two people is more costly than reinforcing an already existing link. 
Clearly, several systems are characterized by this mechanism, where an extra energy is initially required to overcome a threshold, or by the opposite one, where repeated occupations are energetically suppressed due to saturation or aging.
Thus, even if we derive the statistics in the context of networks, they have a wider and more abstract range of application.

A network or graph is a set of $N$ vertices connected by $L$ links or edges.
It is characterized by local properties such as the degree $k_i\equiv\sum_j a_{ij}$ (number of links emanating from vertex $i$, where $a_{ij}=1$ if a link exists between $i$ and $j$, and $a_{ij}=0$ otherwise), and by higher--order correlations, such as the dependence on $k_i$ of the average degree $k^{nn}_i\equiv\sum_j a_{ij}k_j/k_i$ of $i$'s neighbors and the clustering coefficient $c_i\equiv\sum_{jk} a_{ij}a_{jk}a_{ki}/k_i(k_i-1)$.
In real unweighted networks, patterns that were first interpreted as nontrivial \cite{pastor_internet,barabba_clustering} are now  understood as mere effects of the lower--level  graph structure. 
For instance, in a random network where only the degree sequence $\{k_i\}_{i=1}^N$ is specified (the \emph{configuration model}), the probability that the vertices $i$ and $j$ are connected was expected \cite{chunglu} to be 
\begin{equation}
p_{ij}=x_i x_j
\label{eq_CL}
\end{equation}
where $x_i= k_i/\sqrt{2L}$ and $L=\sum_i k_i/2$. This implies that $\langle k_i\rangle=\sum_j p_{ij}=k_i$ \cite{chunglu}, and that $\langle k_i^{nn}\rangle$ and $\langle c_i\rangle$ are independent of $k_i$ (where $\langle \cdots\rangle$ denotes an ensemble average).
Deviations from these flat behaviors were interpreted as a signature of higher--order correlations \cite{pastor_internet,barabba_clustering}. 
However it was later shown that, even for random networks with specified degrees, $k^{nn}_i$ and $c_i$ decrease with $k_i$ \cite{maslov,newman_origin}.  
Indeed, eq.(\ref{eq_CL}) is not the correct probability for large $k_i k_j$, since in this case $p_{ij}>1$, corresponding to undesired multiple edges \cite{newman_origin}. The constraint $p_{ij}<1$ can be enforced by fixing a \emph{structural cut--off} $k_{max}\sim \sqrt{N}$ on the maximum degree \cite{katanza}. However this fails to reproduce real networks, such as the Internet\cite{maslov}, where $k_{max}$ far exceeds this value.
Thus the local properties alone unavoidably determine higher--order `structural correlations' \cite{maslov,newman_origin}. 

Structural correlations can be studied analytically using exponential random graphs \cite{newman_expo}, representing the ensemble of maximally random networks with specified properties $\{\pi_l\}_l$, each governed by a control parameter $\theta_l$. A graph $G$ in the ensemble is assigned the probability $P(G)=e^{-H(G)}/Z$, where 
$H(G)\equiv\sum_l \theta_l\pi_l(G)$
is the graph Hamiltonian and $Z\equiv \sum_G \exp[{-H(G)}]$ is the partition function\cite{newman_expo}. 
Any unweighted graph $G$ is fully specified by its adjacency matrix $A$, with entries $\{a_{ij}\}$. Thus for maximally random graphs with specified degrees \cite{newman_origin} 
$H(A)=\sum_{i=1}^N\alpha_{i}k_{i}=\sum_{i<j}a_{ij}(\alpha_i+\alpha_j)$
and $P(A)=
\prod_{i<j}\frac{e^{-(\alpha_{i}+\alpha_j)a_{ij}}}{1+e^{-(\alpha_{i}+\alpha_j)}}
=\prod_{i<j}p_{ij}^{a_{ij}}(1-p_{ij})^{1-a_{ij}}$, 
where $p_{ij}$ is the probability that $i$ and $j$ are linked:
\begin{equation}
p_{ij}=\frac{e^{-(\alpha_i+\alpha_j)}}{1+e^{-(\alpha_i+\alpha_j)}}=\frac{x_ix_j}{1+x_ix_j}
\label{eq_fermi}
\end{equation}
where $x_i\equiv e^{-\alpha_i}$ is no longer a function of $k_i$ alone. 
The above Fermi--Dirac distribution is the correct null form for $p_{ij}$ \cite{newman_origin}. Since $p_{ij}$ now does not depend only on $k_i$ and $k_j$, higher--order effects are generated even if only local properties are fixed: unlike eq.(\ref{eq_CL}), eq.(\ref{eq_fermi}) correctly predicts that $\langle k^{nn}_i\rangle$ and $\langle c_i\rangle$ decrease with $\langle k_i\rangle$ \cite{newman_origin}. 
Thus purely uncorrelated unweighted networks do not exist.

While the unweighted case is well understood, weighted networks are more controversial. 
On one hand, since structural correlations are due to the `fermionic' constraint disallowing multiple edges \cite{newman_origin}, they are unexpected for weighted graphs, where large weights $w_{ij}$ (equivalent to multiple edges \cite{newman_weighted}) are allowed. 
In particular, if  $s_i\equiv\sum_j w_{ij}$ denotes the \emph{strength} of vertex $i$, random weighted networks with specified strength sequence $\{s_i\}_{i=1}^N$ (we denote this null model as model 3) are expected \cite{weightedconfiguration,serrano2} to follow a weighted version of eq.(\ref{eq_CL}):
\begin{equation}\label{eq_wyy}
\langle w_{ij}\rangle= y_i y_j
\end{equation}
where $y_i=s_i/\sqrt{s_{tot}}$ and $s_{tot}\equiv \sum_i s_i$. This restores the expected degree--independent behavior for the weighted analogues of $k^{nn}_i$ and $c_i$, defined as $k^w_{i}\equiv \sum_j w_{ij}k_j/s_i$ (weighted average nearest neighbor degree, or \emph{affinity}) and $c^w_i\equiv\sum_{j,k}(w_{ij}+w_{ik})a_{ij}a_{ik}a_{jk}/[2s_i(k_i-1)]$ (\emph{weighted clustering coefficient}) respectively \cite{vespy_weighted,vespy_archi}.
On the other hand, theoretical results \cite{newman_expo,ginestra_entropy} (that we confirm and extend later on) indicate that $\langle w_{ij}\rangle$ has a different form, even if the effects on network properties have never been studied. 
Another indicator of correlations is the \emph{disparity} $Y_i\equiv \sum_j w_{ij}^2/s_i^2$ \cite{vespy_archi,barabla}. 
It is expected that $Y_i\approx 1/k_i$ if weights are equally distributed among $i$'s neighbors, and that a larger value of the latter signals an excess concentration of weight in one or more links \cite{vespy_archi,barabla}.
Similarly, the \emph{modularity} \cite{newman_weighted,guimera} (measuring whether the network is partitioned into communities) is expected to vanish for random weighted networks. However, the null behavior of both properties has never been studied systematically.
The nonlinear dependence of $s_i$ on $k_i$ is interpreted as another indicator of correlations \cite{vespy_weighted,vespy_archi}, since if the topology is kept fixed and the weights are globally reshuffled on it \cite{vespy_weighted,vespy_archi}, then $\langle w_{ij}\rangle= \overline{w}a_{ij}$ (where $\overline{w}$ is the average non--zero weight in the network), implying $\langle s_i\rangle=\overline{w}k_i$. However, in this different null model (model 1) $\langle k^w_{i}\rangle$ and $\langle c^w_i\rangle$ equal their unweighted counterparts ($k^{nn}_{i}$ and $c_i$), and thus inherit any purely topological correlation \cite{vespy_weighted}. 
One can partly remove these correlations by globally reshuffling the weights and simultaneously randomizing the topology in a degree--preserving way \cite{colizza2} (model 2). However, the unweighted structural correlations discussed above will still remain. 
The situation becomes even more intricate when both strengths and degrees are prescribed (model 4) \cite{serrano_weighted,manna}. 
This case is difficult to inspect without further assumptions. 
A first interesting result \cite{serrano_weighted} is that it is impossible to decouple purely topological and weighted quantities to obtain completely independent local properties. 
However the assumption of factorized marginal probabilities, leading to an expression analogous to eq.(\ref{eq_wyy}) where $y_i\propto s_i/k_i$, was made  \cite{serrano_weighted}. 
In what follows we go one step further and show that such constraints  represent only a part of the problem. 
We find that the full structural correlations are remarkably stronger, and described in the most general case by mixed Bose-Fermi statistics.

We look for the analytical solutions of the four null models in terms of the probability $q_{ij}(w)$ that $i$ and $j$ are joined by a link of weight $w$ (including $w=0$ when no link is there). The probability $P(W)$ of a graph with weight matrix $W$ (having entries $w_{ij}\ge 0$) is
\begin{equation}
P(W)=\prod_{i<j}q_{ij}(w_{ij})
\label{eq_prodw}
\end{equation}
Without loss of generality we assume integer weights, as in standard approaches \cite{newman_weighted,serrano2,serrano_weighted}. Then $\sum_{w=0}^{w_*} q_{ij}(w)=1$ $\forall i,j$, where $w_{*}$ is the maximum allowed weight. 
The entries of the adjacency matrix are $a_{ij}\equiv\Theta(w_{ij})$, where $\Theta(x)$ is the Heaviside function. 
The probability $p_{ij}$ that $i$ and $j$ are connected by a link, irrespective of the weight of the latter, is 
$p_{ij}=\sum_{w>0}q_{ij}(w)=1-q_{ij}(0)$.
All expectation values are completely specified by $q_{ij}(w)$:
$$\langle k_i\rangle=\sum_j p_{ij}=N-\sum_j q_{ij}(0)$$
$$\langle w_{ij}\rangle=\sum_{w>0} wq_{ij}(w)
\Rightarrow\langle s_i\rangle=\sum_j\langle w_{ij}\rangle$$
$$\langle k^w_i\rangle=\frac{\sum_j \langle w_{ij} k_j\rangle}{\langle s_i\rangle}=
\frac{\sum_j \langle w_{ij}\rangle(\langle k_j\rangle+1-p_{ij})}{\langle s_i\rangle}$$
$$\langle c^w_i\rangle=\frac{\sum_{jk}\langle (w_{ij}+w_{ik})a_{ij}a_{ik}a_{jk}\rangle}{2\langle s_i(k_i-1)\rangle}=
\frac{\sum_{jk}\langle w_{ij}\rangle p_{ik}p_{jk}}{\langle s_i\rangle\langle k_i\rangle-\sum_j \langle w_{ij}\rangle p_{ij}}$$
$$\langle Y_i\rangle=\frac{\sum_j\langle w_{ij}^2\rangle}
{(\sum_k\langle w_{ik}\rangle)^2}=\frac{\sum_j\sum_{w>0} w^2 q_{ij}(w)}
{[\sum_k\sum_{w>0} wq_{ik}(w)]^2}$$
We shall also consider the modularity later on. 
We now reformulate models 1-4 as exponential random graphs. 
In model 1, the whole topology (each entry of $a_{ij}$) is fixed. The only constraint on $w_{ij}$ is then $\Theta(w_{ij})=a_{ij}$:
\begin{equation}
H_1(W)=\sum_{i<j}\alpha_{ij}a_{ij}=\sum_{i<j}\alpha_{ij}\Theta(w_{ij})
\label{eq_H1}
\end{equation}
In models 2-4 the constraints are $\{k_i\}_{i=1}^N$ and/or $\{s_i\}_{i=1}^N$:
\begin{eqnarray}
H_2(W)&=&\sum_{i}\alpha_{i}k_{i}=\sum_{i<j}(\alpha_{i}+\alpha_{j})\Theta(w_{ij})\label{eq_H12}\\
H_3(W)&=&\sum_i\beta_i s_i=\sum_{i<j} (\beta_i+\beta_j)w_{ij}\label{eq_H2}\\
H_4(W)&=&\sum_{i<j} [(\alpha_i +\alpha_j)\Theta(w_{ij})+(\beta_i+\beta_j)w_{ij}]
\label{eq_H3}
\end{eqnarray}
We note that the above models are all particular cases of
\begin{equation}
H(W)=\sum_{i<j} \left[\alpha_{ij}\Theta(w_{ij})+\beta_{ij}w_{ij}\right]
\label{eq_HW}
\end{equation}
The corresponding $P(W)$ can be expressed  as follows:
$$
P(W)=\frac{e^{-H(W)}}{\sum_{W'}e^{-H(W')}}
=\prod_{i<j}\frac{e^{-\alpha_{ij}\Theta(w_{ij})-\beta_{ij}w_{ij}}}
{1+e^{-\alpha_{ij}}\sum_{w'_{ij}=1}^{w_{*}}e^{-\beta_{ij}w'_{ij}}}
$$
Thus, by comparison with eq.(\ref{eq_prodw}), we find analytically
\begin{equation}
q_{ij}(w)=\frac{x_{ij}^{\Theta(w)}y_{ij}^{w}}
{1+x_{ij}\sum_{w'=1}^{w_{*}} y_{ij}^{w'}}
\label{eq_q}
\end{equation}
where we have set $x_{ij}\equiv e^{-\alpha_{ij}}$ and $y_{ij}\equiv e^{-\beta_{ij}}$.
The above class of generalized statistics, interpolating between the Fermi-Dirac ($y_{ij}=1$, or $w_*=1$) and Bose-Einstein ($x_{ij}=1$ and $w_*=+\infty$) ones, is our main result.
It applies to any system described by eq.(\ref{eq_HW}), and represents the probability that its states are populated $w$ times. 
Even if multiple occupations are allowed (which is a property of bosons), the first occupation is necessarily binary (which is a property of fermions). Depending on the sign of $\alpha_{ij}$, the first occupation (whose energy is $\alpha_{ij}+\beta_{ij}$) is either favoured or suppressed with respect to all other occupations, whose energies are $\beta_{ij}$.

We now turn to the four models separately. Models 1 ($y_{ij}=1$) and 2 (for which additionally $x_{ij}=x_i x_j$) yield
$$
q_{ij}(w)=\frac{x_{ij}^{\Theta(w)}}{1+w_{*}x_{ij}};
\qquad p_{ij}=\frac{w_{*}x_{ij}}{1+w_{*}x_{ij}}
$$
Since $\overline{w}=\sum_{w=1}^{w_{*}}w/w_{*}$ and $\overline{w^2}=\sum_{w=1}^{w_{*}}w^2/w_{*}$, we have
\begin{eqnarray}
&\langle k_i\rangle=\sum_j \frac{w_{*}x_{ij}}{1+w_{*}x_{ij}}&
\label{eq_k1}\\
&\langle w_{ij}\rangle=\frac{x_{ij}}{1+w_{*}x_{ij}}\sum_{w=1}^{w_{*}}w
=\overline{w}p_{ij}\Rightarrow\langle s_i\rangle=\overline{w}\langle k_i\rangle&\label{eq_s1}\\
&\langle k^w_i\rangle=\frac{\overline{w}\sum_j p_{ij}(\langle k_j\rangle+1-p_{ij})}{\overline{w}\langle k_i\rangle}=\langle k^{nn}_i\rangle&\label{eq_kw1}\\
&\langle c^w_i\rangle=\frac{\overline{w}\sum_{jk}p_{ij}p_{ik}p_{jk}}{\overline{w}\langle k_i(k_i-1)\rangle}=\langle c_i\rangle&\label{eq_cw1}\\
&\langle Y_i\rangle=\frac{\overline{w^2}\sum_j p_{ij}}
{\overline{w}^2[\sum_j p_{ij}]^2}=\frac{\overline{w^2}}
{\overline{w}^2 \langle k_i\rangle}\label{eq_y1}&
\end{eqnarray}
Equations (\ref{eq_s1}-\ref{eq_cw1}) show that $\langle s_i\rangle\propto k_i$ and that $\langle k^w_i\rangle$ and $\langle c^w_i\rangle$ inherit from $\langle k^{nn}_i\rangle$ and $\langle c_i\rangle$ any dependence on $k_i$, thus conveying information only relative to their unweighted counterparts. 
Moreover, eq.(\ref{eq_y1}) implies $\langle Y_i\rangle\gg 1/\langle k_i\rangle$, since  $\overline{w^2}/\overline{w}^2\gg 1$ for real networks with broadly distributed weights\cite{vespy_weighted,vespy_archi}. However, this reflects the overall weight distribution and does not indicate a local weight imbalance, as usually interpreted \cite{vespy_archi,barabla}. 
These problems arise due to purely fermionic correlations.

\begin{figure}[]	
\includegraphics[width=.5\textwidth]{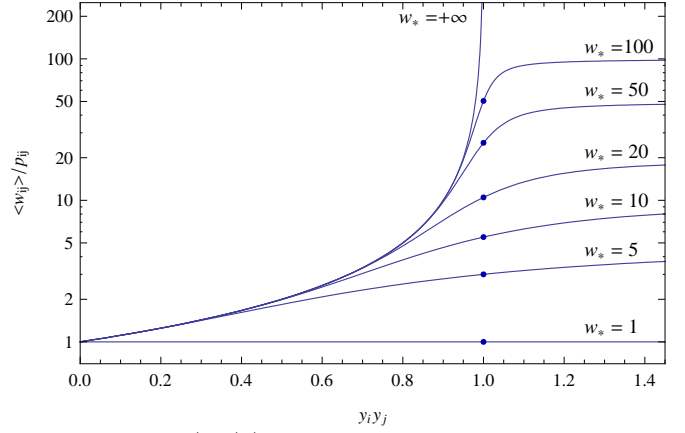}
\caption[]{\small Ratio $\langle w_{ij}\rangle/p_{ij}$ as a function of $y_i y_j$ for models 3 and 4. The values of $\overline{w}$, obtained for $y_i y_j=1$ (models 1 and 2), are highlighted as larger points. 
}
\label{fig_w2}
\end{figure}
\begin{figure}[]	
\includegraphics[width=.5\textwidth]{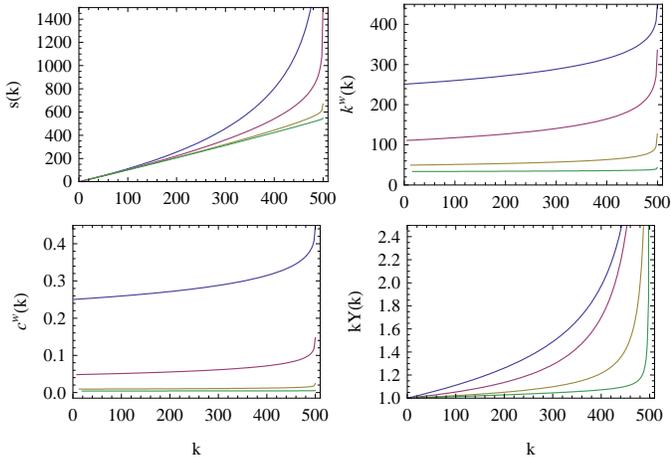}
\caption[]{\small Analytical results for random networks with strength sequence generated by the distribution $\rho(y)\propto y^{-\gamma}$ with (from top to bottom) $\gamma=1,2,3,4$. All networks have the same link density and $N=10000$ vertices.}
\label{fig_m2}
\end{figure}

We now consider model 3 ($x_{ij}=1$, $y_{ij}=y_i y_j$):
$$
q_{ij}(w)=\frac{(y_iy_j)^w(1-y_i y_j)}{1-(y_i y_j)^{w_{*}+1}};\quad 
p_{ij}=\frac{y_i y_j-(y_i y_j)^{w_{*}+1}}{1-(y_i y_j)^{w_{*}+1}}
$$
All the expected properties can again be computed analytically. Their behavior is well revealed by the ratio
\begin{equation}\label{eq:ratio}
\frac{\langle w_{ij}\rangle}{p_{ij}}=
\frac{1}{1-y_i y_j}-\frac{w_* (y_i y_j)^{w_*}}{1-(y_i y_j)^{w_*}}
\end{equation}
which is plotted in fig.\ref{fig_w2}. 
Note that $\langle s_i\rangle=\overline{w}\langle k_i\rangle$ is no longer a correct prediction, since the expectation $\langle w_{ij}\rangle/p_{ij}=\overline{w}$ is never realised.
Similarly, eq.(\ref{eq_wyy}) does not hold. 
All quantities can be calculated for any value of $w_*$. For brevity, we only report the case $w_{*}=+\infty$:
\begin{eqnarray}
&\langle k_i\rangle=\sum_j y_i y_j= N\overline{y} y_i =y_i \sqrt{\langle 2L\rangle}&\label{eq:ky}\\
&\langle w_{ij}\rangle=\frac{y_i y_j}{1-y_iy_j}
\Rightarrow\langle s_i\rangle=\sum_j\frac{y_i y_j}{1-y_iy_j}&\label{eq:w2}\\
&\langle k^w_i\rangle=\sum_j\frac{y_j(1+ N\overline{y}y_j-y_iy_j)}{1-y_iy_j}/\sum_j\frac{y_j}{1-y_iy_j}&\\
&\langle c^w_i\rangle=\sum_{j} \frac{y_j^2 \sum_k y_k^2}{1-y_i y_j}/
\sum_j\frac{y_j(N\overline{y}-y_j)}{1-y_i y_j}&\\
&\langle Y_i\rangle=\sum_j\frac{y_i y_j(1+y_iy_j)}{(1-y_iy_j)^2}/[\sum_j\frac{y_i y_j}{1-y_iy_j}]^2&
\end{eqnarray}
where $\overline{y}\equiv\sum_i y_i/N$. 
As clear from eq.(\ref{eq:ky}), $y_i\propto \langle k_i\rangle$ and therefore all the above quantities display a nontrivial dependence on the degree. 
For instance, we can study scale--free networks by considering a power--law distribution $\rho(y)\propto y^{-\gamma}$ for the $y_i$'s (implying $P(k)\propto k^{-\gamma}$), and approximating the discrete sums with (analytically solvable) integrals: $\sum_j\to N\int dy\rho(y)$.
The resulting curves, shown in fig.\ref{fig_m2}, strongly contradict the expectations \cite{vespy_weighted,vespy_archi,barabla} that one should observe $s_i\propto k_i$, and all other curves flat. 
While fermionic correlations yield disassortative trends, bosonic correlations generate assortative patterns. These constraints have a deeper origin than those studied in \cite{serrano_weighted}, where the unavoidable dependence of weights on connectivity was considered. Here we find that even if only the strengths (and not the degrees) are fixed, then $\langle w\rangle$ does not factorize. 

Finally, in model 4 ($x_{ij}=x_i x_j$, $y_{ij}=y_iy_j$) the ratio $\langle w_{ij}\rangle/p_{ij}$ is still given by eq.(\ref{eq:ratio}). 
Now $q_{ij}(w)$ reads
\begin{equation}
q_{ij}(w)=\frac{(x_ix_j)^{\Theta(w)}(y_iy_j)^{w}(1-y_iy_j)}
{1-y_iy_j+x_ix_jy_iy_j- x_ix_j(y_iy_j)^{1+w_*} }
\label{eq_q3}
\end{equation}
Here one sees that, even if $x_i$ and $y_i$ are chosen as statistically independent, the resulting weighted and purely topological quantities are not independent of each other. This is the effect studied in \cite{serrano_weighted} that we automatically recover here. However, eq.(\ref{eq_q3}) also takes into account both bosonic and fermionic constraints. Therefore, unlike \cite{serrano_weighted}, here we do not need to restrict ourselves to sparse networks. We conclude that for models 3 and 4 the available weighted measures are uninformative, either in an absolute or in a relative sense. Thus a systematic redefinition of weighted network properties is necessary.

Structural correlations also affect the modularity of a partition of a network into communities, defined as
$$
Q\equiv\sum_{i<j}\left[\frac{a_{ij}}{2L}-\frac{k_i k_j}{(2L)^2}\right]c_{ij},\quad
Q_{w}\equiv\sum_{i<j}\left[\frac{w_{ij}}{s_{tot}}-\frac{s_i s_j}{s_{tot}^2}\right]c_{ij} 
$$
in the unweighted and weighted case respectively \cite{newman_weighted}, where $c_{ij}=1$ if $i$ and $j$ belong to the same community, and $c_{ij}=0$ otherwise.
For a non-modular network with only local constraints, 
one expects $\langle Q\rangle=0$ since the differences in the square 
brackets are expected to vanish according to eqs.(\ref{eq_CL}) 
and (\ref{eq_wyy}). However, we have shown that these 
expectations are wrong. Thus $\langle Q\rangle\ne 0$ even for 
random graphs, which means that the modularity of real networks 
is unavoidably biased and does not entirely represent a signature 
of community structure. Interestingly, the reverse operation, i.e. 
randomizing an unweighted network keeping both the modularity 
and the degree sequence fixed, has been shown \cite{guimera} 
to reproduce most of the observed degree-degree correlations. 
 
Our formalism treats null models in a unified fashion, but it clearly cannot indicate \emph{a priori} the most appropriate null model for a specific network. 
Nonetheless, the identification of the Hamiltonian corresponding to each model allows deep insights into network structure.
For instance, we can interpret in a new light the results \cite{serrano_weighted,manna} showing that some real networks, such as the US airport network and the World Trade Web, remain almost unchanged after  randomizations that preserve both strengths and degrees.
Indeed, for such networks  the establishment of a link for the first time requires an extra cost (a transportation channel and/or a trade agreement), while on already existing links any further interaction is facilitated. 
In general, for these and other systems (including our initial example of social networks) eq.(\ref{eq_HW}) may be already a good model, not simply a null one. If this is the case, the Bose-Fermi statistics in eq.(\ref{eq_q}) will naturally describe such real systems.

\end{document}